\documentclass[conference]{IEEEtran}
\IEEEoverridecommandlockouts
\usepackage{cite}
\usepackage{url} % 基础URL格式化

\usepackage{hyperref} % 移除了禁用选项
\hypersetup{
    colorlinks  = true,   % 启用彩色链接
    linkcolor   = black,   % 内部链接颜色
    citecolor   = black,   % 引用链接颜色
    urlcolor    = black,   % URL链接颜色
    pdfusetitle = true,   % 允许生成PDF元数据
    pdfauthor   = {},     % 作者信息（按需填写）
    pdftitle    = {},     % 标题信息（按需填写）
}

\usepackage{amsmath,amssymb,amsfonts}
\usepackage{bm}

\usepackage{algorithmic}
\usepackage[ruled,vlined,linesnumbered]{algorithm2e}

\usepackage{graphicx}
\usepackage{subcaption}

\renewcommand{\thesubsection}{\thesection.\Alph{subsection}}


\begin{document}

\title{Difference Imaging-Based Parking Lot Surveillance in Multi-RIS-Aided Collaborative ISAC System\\
}

% 作者信息定义
\author{
\IEEEauthorblockN{Zhengze Ji\textsuperscript{1}, 
Yixuan Huang\textsuperscript{2}, 
Zhixin Chen\textsuperscript{1}, 
Jie Yang\textsuperscript{3,4}, 
and Shi Jin\textsuperscript{2,4}}
\IEEEauthorblockA{\textsuperscript{1}Chien-Shiung Wu College, Southeast University, Nanjing, China}
\IEEEauthorblockA{\textsuperscript{2}National Mobile Communications Research Laboratory, Southeast University, Nanjing, China}
\IEEEauthorblockA{\textsuperscript{3}Key Laboratory of Measurement and Control of Complex Systems of Engineering, Ministry of Education}
\IEEEauthorblockA{\textsuperscript{4}Frontiers Science Center for Mobile Information Communication and Security, Southeast University, Nanjing, China}
\IEEEauthorblockA{Email: \{jizhengze, huangyx, chenzhixin, yangjie, jinshi\}@seu.edu.cn}
}

\maketitle

\begin{abstract}
Parking lot surveillance with integrated sensing and communication (ISAC) system is one of the potential application scenarios defined by 3rd Generation Partnership Project (3GPP). Traditional surveillance systems using cameras or magnetic sensors face limitations such as light dependence, high costs, and constrained scalability. Wireless sensing with reconfigurable intelligent surfaces (RISs) has the ability to address the above limitations due to its light independence and lower deployment overhead. In this study, we propose a difference imaging-based multi-RIS-aided collaborative ISAC system to achieve parking lot surveillance. In a parking lot, the presence of vehicles induces impacts on wireless environments due to scattering characteristic variation. By delineating the parking lot into a two-dimensional image with several grid units, the proposed system can capture the variation of their scattering coefficients in free and occupied states. The variation between these two states is sparse, which can be captured through compressed sensing (CS)-based imaging algorithms. Additionally, we collaboratively employ multiple RISs to enable higher surveillance performance. Experimental results demonstrate that our method can achieve high-accuracy parking occupancy detection, and the employment of collaborative RISs further enhances the detection rate.
\end{abstract}

\section{Introduction}
With accelerating urbanization and rising vehicle ownership, real-time parking lot surveillance is critical for improving efficiency and reducing congestion \cite{Amato2016Parking}. Conventional camera-based systems struggle in low-light conditions \cite{Grodi2016Smart}, while magnetic sensors incur high installation costs \cite{Wang2009Vehicle}, creating demand for scalable alternatives.

Integrated sensing and communication (ISAC) offers a promising solution by enabling simultaneous wireless data transfer and environmental sensing \cite{Cheng2024Intelligent}. Recognized by 3GPP as ideal for smart city applications \cite{3GPPTR22837,Li2023ISAC}, ISAC provides cost-effective scalability and environment-adaptive sensing \cite{Tan2021Integrated}. Reconfigurable intelligent surfaces (RISs) further enhance ISAC through dynamic channel manipulation, enabling innovations like near-field imaging \cite{Jiang2024NearField} and localization \cite{Huang2023Localization}.

For parking surveillance, RIS-assisted imaging outperforms localization methods by directly estimating target scattering information \cite{Huang2023Localization}. However, current RIS-assisted imaging technologies still face limitations. Existing studies predominantly employ a single RIS, resulting in a restricted imaging aperture and suboptimal accuracy \cite{Taha2023Sensing}, \cite{Hu2022MetaSketch}, \cite{Zhang2022Sensing}. In a later study, it has been shown that improvements in imaging performance can be obtained when employing two RISs \cite{Zhao2023Intelligent}. This inspires us to use multi-RIS collaboration to create large imaging apertures and improve imaging performance \cite{Huang2024RISImaging}.

Current RIS-aided imaging methodologies consist mainly of two categories: Fourier transform (FT) \cite{Zhang2012Sparse} and compressed sensing (CS) \cite{Fang2014Fast}. Compared to FT-based algorithms, which require extensive measurements for image reconstruction, CS-based algorithms require significantly fewer measurements, leading to substantially reduced data acquisition time and thereby enabling real-time surveillance. Moreover, CS-based algorithms can exploit the inherent sparsity of the imaging area to achieve high-accuracy results \cite{Huang2024Fourier}. However, the parking lot images, which represent the scattering characteristics of the parking lot, are typically not sparse. This is because the scattering coefficients of the concrete floor and the cars are not zero. To address this problem, we propose to harness the sparsity lying in the parking lot difference image. Specifically, the difference image is defined as the difference of parking lot images with and without vehicles, which is inherently sparse. Therefore, CS-based imaging algorithms can be employed for detecting the parking space status.

The main contributions of this paper are summarized as follows.
\begin{itemize}
\item We propose a parking lot surveillance method by estimating its difference image with the aid of RIS.
\item We collaboratively deploy multiple RISs, significantly improving the accuracy of parking space status detection to 100\% under certain conditions.
\end{itemize}

The remainder of this paper is organized as follows. We introduce the system configuration and channel model for parking lot surveillance in \hyperref[sec:ii]{Sec.~\ref*{sec:ii}}. In \hyperref[sec:iii]{Sec.~\ref*{sec:iii}}, we specify the problem we need to address. \hyperref[sec:iv]{Sec.~\ref*{sec:iv}} presents the multi-RIS-aided surveillance method. In \hyperref[sec:v]{Sec.~\ref*{sec:v}}, the simulation results are demonstrated. Finally, we draw conclusions and outline future work in \hyperref[sec:vi]{Sec.~\ref*{sec:vi}}.

\section{System Model}\label{sec:ii}

\begin{figure}[htbp]
\centerline{\includegraphics[width=\linewidth]{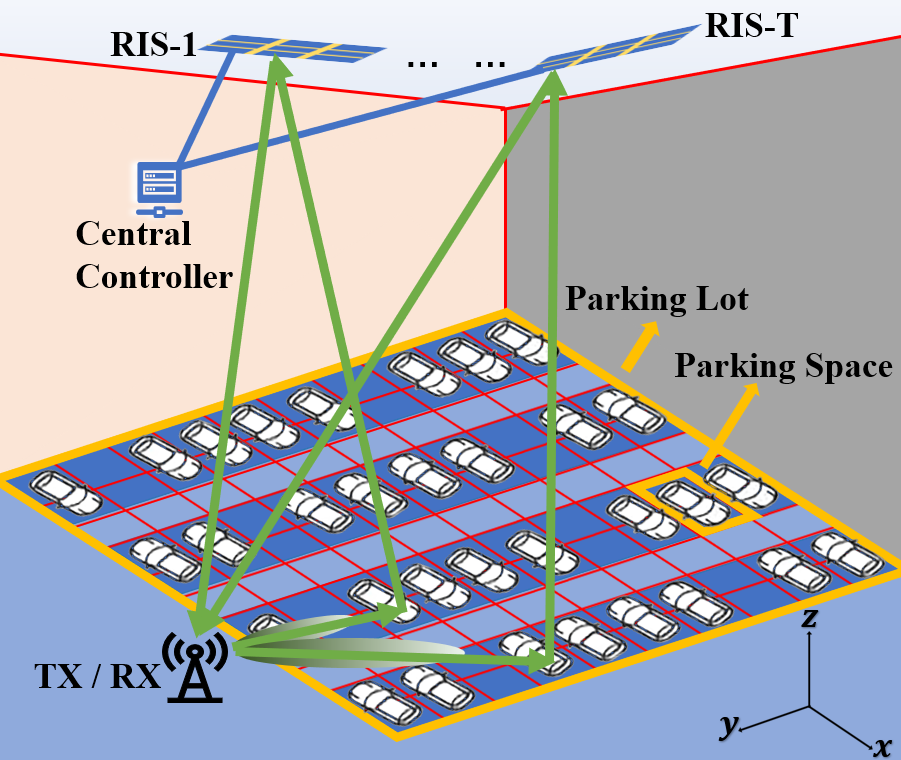}}
\caption{Illustration of the multi-RIS-aided parking lot surveillance ISAC system.}
\label{fig:htbp} % 修改标签为 fig:htbp
\end{figure}

In this study, we consider a multi-RIS-aided parking lot surveillance ISAC system in the three-dimensional (3D) space \(\mathbb{R}^3 = \{\, [x, y, z]^\text{T}\mid x, y, z \in \mathbb{R} \,\}\), as depicted in Fig.~\ref{fig:htbp}. The entire parking lot, which is the region of interest (ROI) of the proposed imaging function, is modeled as a two-dimensional (2D) plane for simplicity.\footnote{ Future work will extend it to a 3D scenario.} It is divided into \( Q \) spatial grid units, with every \( C \) units covering an individual parking space. The scattering coefficient of the \( q \)-th spatial grid unit in the ROI is denoted as \( \sigma_q \). \(\bm{\sigma} = \begin{bmatrix} \sigma_1,\ldots,\sigma_Q \end{bmatrix}^\text{T} \in \mathbb{Q}^{Q \times 1} \) is a column vector formed by scattering coefficients of all units. We have a full duplex base station that contains both the transmitter (TX) and receiver (RX), which are equipped with \( N_{\text{tx}} \) and \( N_{\text{rx}} \) antennas, respectively. Moreover, the sensing process is performed during the communication process and the implementation of the sensing function has little impact on the user's communication. Additionally, \( T \) RISs are deployed at different positions on the ceiling of the parking lot, with the height of \(\hbar\). RIS arrays are parallel to the ground, forming a distributed multi-RIS-aided sensing system. The \( t \)-th RIS consists of \( M_t \) tunable units and the RIS phase configurations are controlled by a central controller.

\subsection{Received Signal Model}  

In this study, we assume that the TX transmits \( K \) sensing signals during a single imaging session of the parking lot. During this period, the RIS phase configurations are randomly generated in each signal interval. Denoting \( p \) as the transmitted signal, the received signal at the RX in the \( k \)-th symbol interval is given as \cite{Huang2024RISImaging}
\begin{equation}
  \mathbf{r}_{k} = \left( \mathbf{h}^{\text{LOS}} + \mathbf{h}^{\text{ROI}} + \mathbf{h}_{k}^{\text{RIS}} + \mathbf{h}_{k}^{\text{Ima}} + \mathbf{h}_{k}^{\text{other}} \right) p + \tilde{\mathbf{n}}_{k},
  \label{eq:1}
\end{equation}  
where \(\mathbf{h}^{\text{LOS}}\), \(\mathbf{h}^{\text{ROI}}\), \(\mathbf{h}^{\text{RIS}}_{k}\), and \(\mathbf{h}^{\text{Ima}}_{k}\) represent the channel responses of the direct TX-RX path, TX-ROI-RX path, TX-RIS-RX path, and TX-ROI-RIS-RX path, respectively. \(\mathbf{h}_{k}^{\text{other}}\) accounts for residual multipath effects caused by random scatterers and multi-bounce reflections. \(\tilde{\mathbf{n}}_{k}\) represents additive noise.  

The channel component \(\mathbf{h}_{k}^{\text{Ima}} \), shaped by interactions with the ROI and RIS, embeds scattering features of the ROI while generating distinct channel responses through RIS phase variations. This component can be isolated from other multipath signals via channel estimation \cite{Wang2023Reconfigurable} and be employed for parking lot imaging.   

\subsection{Channel Model of the TX-ROI-RIS-RX Path}

For the TX-ROI-RIS-RX path, the subpath starting from the \( i \)-th transmitting antenna during the \( k \)-th emission, scattered by the \( q \)-th spatial grid unit and the \( m \)-th unit of the \( t \)-th RIS, ultimately reaches the \( j \)-th receiving antenna can be expressed as \cite{Huang2024RISImaging}
\begin{align}
    h_{i,k,q,t,m,j} 
    &= g 
    \frac{e^{-j2\pi\frac{d_{\text{TX}_i, \text{v}_q}}{\lambda}}}{\sqrt{4\pi}d_{\text{TX}_i, \text{v}_q}} 
    \times\sigma_q\times 
    \frac{e^{-j2\pi\frac{d_{\text{v}_q, \text{s}_{t,m}}}{\lambda}}}{\sqrt{4\pi}d_{\text{v}_q, \text{s}_{t,m}}}\times 
    e^{-j\omega_{k,t,m}} \nonumber \\
    &\quad \times\frac{e^{-j2\pi\frac{d_{\text{s}_{t,m}, \text{RX}_j}}{\lambda}}}{\sqrt{4\pi}d_{\text{s}_{t,m}, \text{RX}_j}} \label{eq:channel_model},
\end{align}
where 
 \( g \) is a constant determined by the antenna gains of the TX and RX;
 \( \lambda \) is the wavelength; \( e^{-j\omega_{k,t,m}} \) represents the RIS phase configuration of the \( m \)-th unit of the \( t \)-th RIS during the emission of the \( k \)-th sensing signal;
 \( d_{*_\circ, \star_\bullet} \) denotes the distance between the \( \circ \)-th element of \( * \) and the \( \bullet \)-th element of \( \star \), where \( *, \star \in \{\text{TX}, \text{v}, \text{s}_t, \text{RX}\} \) represent the TX, ROI, the \( t \)-th RIS, and the RX, respectively, and \( \circ, \bullet \in \{i, q, m, j\} \).

By summing up all channels scattered by the \( Q \) spatial grid units and the \( M_t \) tunable units of the \( t \)-th RIS, we derive
\begin{equation}
\begin{split}
h_{i,k,t,j} &= \sum_{q=1}^{Q} \sum_{m=1}^{M_t} h_{i,k,q,t,m,j} \\
&= g \, \mathbf{h}_{\text{TX}_i,\text{v}}^\text{T} \, \operatorname{diag}(\boldsymbol{\sigma}) \, \mathbf{H}_{\text{v},s_t} \, \operatorname{diag}(\boldsymbol{\omega}_{k,t}) \, \mathbf{h}_{s_t,\text{RX}_j}
\end{split}
\label{eq:combined},
\end{equation}
where
\(\boldsymbol{\omega}_{k,t} = \begin{bmatrix} \omega_{k,t,1},\ldots,\omega_{k,t,M_t} \end{bmatrix}^\text{T} \in \mathbb{C}^{M_t \times 1}\) represents the phase configuration of the \( t \)-th RIS in the \( k \)-th symbol interval. 
\( \mathbf{h}_{\text{s}_t,\text{RX}_j} \in \mathbb{C}^{M_t \times 1} \), representing the channel from the \( t \)-th RIS to the \( j \)-th receiving antenna, is defined as
\begin{align}
\mathbf{h}_{\text{s}_t,\text{RX}_j} &= 
    \begin{bmatrix}
        \frac{e^{-j2\pi\frac{d_{\text{s}_{t,1}, \text{RX}_j}}{\lambda}}}{\sqrt{4\pi}d_{\text{s}_{t,1}, \text{RX}_j}},\ldots,
        \frac{e^{-j2\pi\frac{d_{\text{s}_{t,M_t}, \text{RX}_j}}{\lambda}}}{\sqrt{4\pi}d_{\text{s}_{t,M_t}, \text{RX}_j}}
    \end{bmatrix}^\text{T}. \label{eq:5}
\end{align}
\( \mathbf{h}_{\text{TX}_i,\text{v}} \in \mathbb{C}^{Q \times 1} \) and \( \mathbf{H}_{\text{v},s_t} \in \mathbb{C}^{Q \times M_t} \) represent the channels from the \( i \)-th transmitting antenna to the ROI and from the ROI to the \( t \)-th RIS, respectively. They possess similar forms to \( \mathbf{h}_{\text{s}_t,\text{RX}_j} \).

\section{Problem Formulation}\label{sec:iii}  

By transmitting and receiving signals, varying the RIS phase configurations, and conducting channel estimation, we can acquire the CSI measurement \( y_{i,k,t,j} \). It corresponds to the \( i \)-th transmitting antenna, the \( k \)-th sensing signal, the \( t \)-th RIS, and the \( j \)-th receiving antenna, given as
\begin{align}
  {y}_{i,k,t,j} = {h}_{i,k,t,j} + {n}_{i,k,t,j},\label{eq:add}
\end{align}
where \({n}_{i,k,t,j}\) is the noise originated from channel estimation. \( K \) CSI measurements like~\eqref{eq:add} with \( K \) different RIS phases can be stacked to obtain \(\mathbf{y}_{i,t,j} = \begin{bmatrix} 
y_{i,1,t,j}, \ldots, y_{i,K,t,j} 
\end{bmatrix}^\text{T} \in \mathbb{C}^{K \times 1}\), which satisfies
\begin{align}
  \mathbf{y}_{i,t,j} = \mathbf{A}_{i,t,j} \, \bm{\sigma} + \mathbf{n}_{i,t,j},\label{eq:9}
\end{align}
where \( \mathbf{n}_{i,t,j} = \begin{bmatrix} {n}_{i,1,t,j},\ldots,{n}_{i,K,t,j} \end{bmatrix}^\text{T} \in \mathbb{C}^{K \times 1} \) is the additive noise. The  \( k \)-th row of the sensing matrix \(\mathbf{A}_{i,t,j}\in C^{K\times Q}\) is \(\mathbf{h}_{i,k,t,j} \), given as
\begin{align}
\mathbf{h}_{i,k,t,j} = g \, \text{diag}\left(\mathbf{h}_{\text{TX}_i,\text{v}}\right) \, \mathbf{H}_{\text{v},\text{s}_t} \, \text{diag}\left(\boldsymbol{\omega}_{k,t}\right) \mathbf{h}_{\text{s}_t,\text{RX}_j}.\label{eq:6}
\end{align}

By stacking the CSI measurements corresponding to the \( N_{\text{tx}} \) transmitting antennas, \( T \) RISs, and \( N_{\text{rx}} \) receiving antennas, the total CSI measurements can be written as \(\mathbf{y} = \begin{bmatrix} \mathbf{y}_{1,1,1}^\text{T},\ldots,\mathbf{y}_{N_{\text{tx}},T,N_{\text{rx}}}^\text{T} \end{bmatrix}^\text{T} \in \mathbb{C}^{N_{\text{A}} \times 1}\), where \( N_{\text{A}} = N_{\text{tx}}\, K\, T\, N_{\text{rx}}\). For the vector \( \mathbf{y}\), it satisfies 
\begin{align}
  \mathbf{y} = \mathbf{A} \, \bm{\sigma} + \mathbf{n},\label{eq:9}
\end{align}
where  \( \mathbf{n} = \begin{bmatrix}
\mathbf{n}_{1,1,1}^\text{T},\ldots,\mathbf{n}_{N_{\text{tx}},T,N_{\text{rx}}}^\text{T} \end{bmatrix}^\text{T}\in \mathbb{C}^{N_{\text{A}} \times 1} \). \(\mathbf{A} = \begin{bmatrix} \mathbf{A}_{1,1,1}^\text{T},\ldots,\mathbf{A}_{N_{\text{tx}},T,N_{\text{rx}}}^\text{T} \end{bmatrix}^\text{T} \in \mathbb{C}^{N_{\text{A}} \times Q}\) is the sensing matrix of the whole sensing system. \(\mathbf{A}\) remains invariant during the sensing process due to the predefined RIS phase configurations and the fixed locations of the TX, RX, RISs, and ROI.

In this study, we need to detect the status of parking spaces with the knowledge of \( \mathbf{y}\) and \(\mathbf{A}\). Equivalently, we need to acquire a parking space status image \(\boldsymbol{\Omega} = [\Omega_1, \ldots, \Omega_Q]^\text{T} \in \{0,1\}\), where \(\Omega_q\) is defined as
\begin{equation}
\Omega_q = 
\begin{cases}
1, & \text{if the } q\text{-th grid unit is occupied} \\
0, & \text{if the } q\text{-th grid unit is free}
\end{cases}.
\end{equation}
Basically, the parking space status image \(\boldsymbol{\Omega}\) can be obtained based on the scattering coefficient vector \( \bm{\sigma} \). However, \( \bm{\sigma} \) exhibits no sparsity, rendering direct application of CS-based algorithms infeasible. As a result, it is difficult to directly derive \( \bm{\sigma} \). In the next section, we will propose a difference imaging algorithm to solve this problem.

\section{Difference Imaging Algorithm}\label{sec:iv}
To address the problem formulated in the above section, a three-stage difference imaging method is proposed, which is enumerated in Algorithm~\ref{alg:parking_monitoring}. We initially acquire reference measurements when no vehicles exist in the parking lot. Then, measurements with vehicles are captured. Finally, we utilize the measurement difference to realize parking lot difference imaging and parking space status detection.

\begin{algorithm}[!t]
  \caption{Difference Imaging Algorithm} 
  \label{alg:parking_monitoring}
  \SetAlgoLined
  \KwIn{
    Vehicle-free measurement $\mathbf{y}^0$,
    occupancy measurement $\mathbf{y}^1$,
    sensing matrix $\mathbf{A}$, and
    sparsity threshold $\tau$.
  }
  
  $\Delta\mathbf{y} \leftarrow \mathbf{y}^1 - \mathbf{y}^0$
  
  Establish model $\Delta\mathbf{y} = \mathbf{A}\Delta\bm{\sigma} + \mathbf{n}^1$
  
  Estimate $\Delta\bm{\sigma}$ using the SP algorithm
  
  Initialize status image $\bm{\Omega} \leftarrow \mathbf{0}$ \;
  \For{each grid unit $q = 1$ \KwTo $Q$}{
    \If{$\Delta\sigma_q > \tau$}{
      Mark as occupied: $\Omega_q \leftarrow 1$ \;
    }
  }
  
  \KwOut{Parking status image $\bm{\Omega} \in \{0,1\}$.}
\end{algorithm}

\subsection{Reference Measurements Acquisition with No Vehicles}\label{sec:iv_a}

Under the condition that the parking lot has no vehicles, there is merely concrete floor in the ROI. The ROI image is denoted as \(\bm{\sigma}^0 = \begin{bmatrix} \sigma^0_1,\ldots,\sigma^0_Q \end{bmatrix}^\text{T}\). During the whole reference measurements acquisition stage, the system performs \( P \) independent measurement acquisition sessions. In each session, the TX emits \( K \) sensing signals while RISs alter their phases. The reference channel measurements derived in the \( p \)-th session is given as
\begin{equation}
    \mathbf{y}^0_p = \mathbf{A} \bm{\sigma}^0 + \mathbf{n}^0_p, 
\end{equation}
where  \( p = \{1,\ldots,P\}\). \( \mathbf{n}^0_p\) is the additive noise.

To suppress measurement noise, the final reference measurement is derived as the average of $\{\mathbf{y}_p^0\}_{p=1}^P$, given as
\begin{equation}
    {\mathbf{y}}^0 = \frac{1}{P} \sum_{p=1}^P \mathbf{y}^0_p = \mathbf{A}\bm{\sigma}^0 + \frac{1}{P}\sum_{p=1}^P \mathbf{n}^0_p.\label{eq:noise}
\end{equation}
As \( P \to \infty \), the energy of the noise term in~\eqref{eq:noise} converges to zero according to the law of large numbers. Thus, the final reference measurement can be approximated as
\begin{equation}
    \mathbf{y}^0 \approx \mathbf{A} \bm{\sigma}^0.
    \label{eq:noise_free_ref}
\end{equation}

\subsection{Measurements Acquisition with Vehicles}

When vehicles are present in the parking lot, the scattering coefficient vector of grid units in the ROI is changed, which is denoted as \(\bm{\sigma}^1 = \begin{bmatrix} \sigma^1_1,\ldots,\sigma^1_Q \end{bmatrix}^\text{T}\). It is worth mentioning that \(\bm{\sigma}^1\) and \(\bm{\sigma}^0\) have the same element value where the status of the corresponding grid unit is free. The different element values only exist when the units are occupied by vehicles. During this stage, the system performs merely one measurement acquisition session with \( K \) sensing signals to ensure real-time imaging and status detection. The channel measurements with vehicles can be given as
\begin{align}
  \mathbf{y}^1 = \mathbf{A} \, \bm{\sigma}^1 + \mathbf{n}^1,\label{eq:8}
\end{align}
where \( \mathbf{n}^1 \in \mathbb{C}^{N_{\text{A}} \times 1} \) is additive Gaussian noise.

Subtracting \( \mathbf{y}^0 \) from \( \mathbf{y}^1 \), we derive
\begin{align}
    \Delta\mathbf{y} &= \mathbf{y}^1 - \mathbf{y}^0 \nonumber \\
                     &\approx \mathbf{A}(\bm{\sigma}^1 - \bm{\sigma}^0) + \mathbf{n}^1 \nonumber \\
                     &= \mathbf{A}\Delta\bm{\sigma} + \mathbf{n}^1 \label{eq:final_form},
\end{align}
where \( \Delta\boldsymbol{\sigma} = \boldsymbol{\sigma}^1 - \boldsymbol{\sigma}^0  = \begin{bmatrix} \Delta\sigma_1,\ldots,\Delta\sigma_Q \end{bmatrix}^\text{T}\), representing the difference image induced by the existence of vehicles.

\subsection{Difference Imaging and Occupancy Detection}

We assume that only a small number of parking spaces are occupied, thus \(\mathrm{\Delta}\bm{\sigma}\) is a sparse vector. Its sparsity can be exploited by employing CS-based algorithms, thereby obtaining the difference image. Even if parking spaces are heavily occupied, it does not matter because our monitoring is not intended for static scenes but rather for dynamic changes. In other words, once our system begins operating, we can obtain measurements of the parking lot at regular intervals. As long as the time intervals are appropriately selected, the changes in the parking lot's status between two measurements will be sparse. In this case, the reference measurement is not the measurement taken with no vehicles, but rather the measurement obtained in the previous interval, thereby satisfying the sparsity condition.

Specifically, the subspace pursuit (SP) algorithm \cite{Becerra2018Sparse} is adopted to recover \(\Delta\boldsymbol{\sigma}\) in this study, which iteratively refines the support set by selecting dominant components while maintaining computational efficiency. This approach achieves a balance between reconstruction accuracy and computational complexity \cite{Dai2009SP}, making it particularly suitable for parking occupancy detection with limited measurements. 

After deriving the sparse vector \(\Delta\bm{\sigma}\), the occupancy status of the parking space can be determined through the following steps. Initially, we decide the status of each grid unit. Let \( \tau_1 \) and \( \tau_2 \) be the thresholds for distinguishing free and occupied status in spatial grid units. For each grid unit \( q \), if the value of \(\Delta\sigma_q\) is between \( \tau_1 \) and \( \tau_2 \), the unit is classified as occupied; otherwise, it is considered free. As a result, \(\boldsymbol{\Omega}\) satisfies
\begin{equation}
  \Omega_q = 
  \begin{cases} 
    1 & \text{if } \tau_1 < \Delta\sigma_q < \tau_2 \\
    0 & \text{otherwise}
  \end{cases}.
  \label{eq:10}
\end{equation}

We then determine the status of each individual parking space based on the status of the grid units. A parking space is marked as occupied when the proportion of its associated spatial grid units with \(\Omega_q = 1\) reaches or exceeds a threshold of \({\eta}\). Otherwise, it is declared unoccupied. Therefore, we realize parking lot surveillance through multi-RIS-aided difference imaging.

\begin{figure}[htbp]
\centerline{\includegraphics[width=\linewidth]{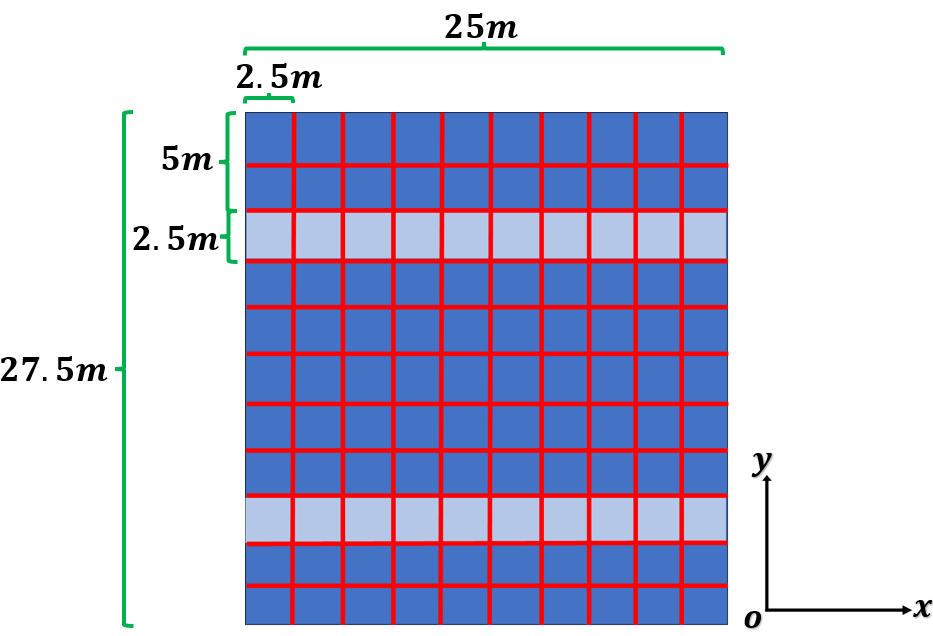}}
\caption{Illustration of the 2D ROI scenario in the simulations.}
\label{fig:htbpn} % 修改标签为 fig:htbpn
\end{figure}

\section{Numerical Results}\label{sec:v}

\subsection{Experiment Setup and Evaluation Metrics}

In the simulation experiment, the ceiling height of the parking lot is \(\hbar = 3 \) meters. The ROI is a rectangular region as depicted in Fig.~\ref{fig:htbpn}. The center of the ROI is the origin of the axes. The dimension along the \(x\) axis of the ROI is 25 meters, divided into 10 parking spaces, each with a width of 2.5 meters. The dimension along the \(y\) axis is 27.5 meters, with a total of 4 rows of parking spaces, each with a length of 5 meters. This means that the size of each parking space is \(2.5\,\text{m} \times 5\,\text{m}\). The spacing between each row is 2.5 meters, which serves as the driving lane. The ROI is divided into \(10 \times 11\) spatial grid units, which means \( Q = 110 \). Each grid unit is with the size of \(2.5\,\text{m} \times 2.5\,\text{m}\), which means that every \( C = 2 \) units covers an individual parking space.

We define that \(\bm{\sigma}^0\) follows a uniform distribution within the interval \([0.45, 0.55]\). For \(\bm{\sigma}^1\), most of the elements are equal to the corresponding elements in \(\bm{\sigma}^0\), while other elements vary due to the existence of vehicles, following a uniform distribution within the interval \([0.85, 0.95]\). The number of varying grid units is twice the vehicle count in the parking lot. We define \( \tau_1 = 0.25 \) and \( \tau_2 = 0.55 \) as the thresholds for distinguishing the status of spatial grid units and \(\eta = 50\% \) as the threshold for distinguishing the status of parking spaces. 

We utilize \( N_{\text{tx}} = 1 \) transmitting antenna and \( N_{\text{rx}} = 1 \) receiving antenna, which are both located at $[-12.5, -13.75, 1]^{\text{T}}$ meters. The TX transmits \( K = 300\) sensing signals during a single imaging session of the parking lot. The signals operate at a frequency of \(3~\text{GHz}\). We utilize \( T = 2 \) RISs to collaboratively achieve ROI imaging, which are located at $[7.5, 0, 3]^{\text{T}}$ meters and $[-7.5, 0, 3]^{\text{T}}$ meters, respectively. Both RISs consist of \(50 \times 50\) tunable units. Each tunable unit has the size of \(0.05\,\text{m} \times 0.05\,\text{m}\).

\begin{figure*}[!t] % 注意星号 *
   \centering
  % 子图 (a)
  \begin{subfigure}[b]{0.3\linewidth}
    \includegraphics[width=\linewidth]{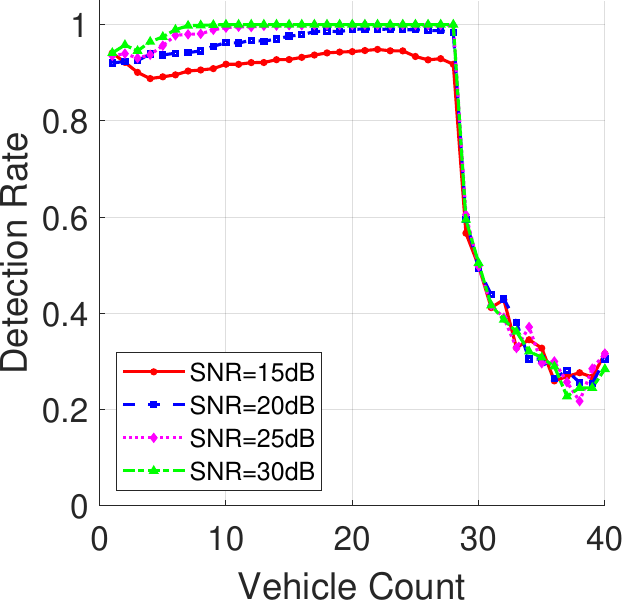}
    \caption{}
    \label{fig:sub_slot}
  \end{subfigure}
  \hfill % 填充水平间距
  % 子图 (b)
  \begin{subfigure}[b]{0.3\linewidth}
    \includegraphics[width=\linewidth]{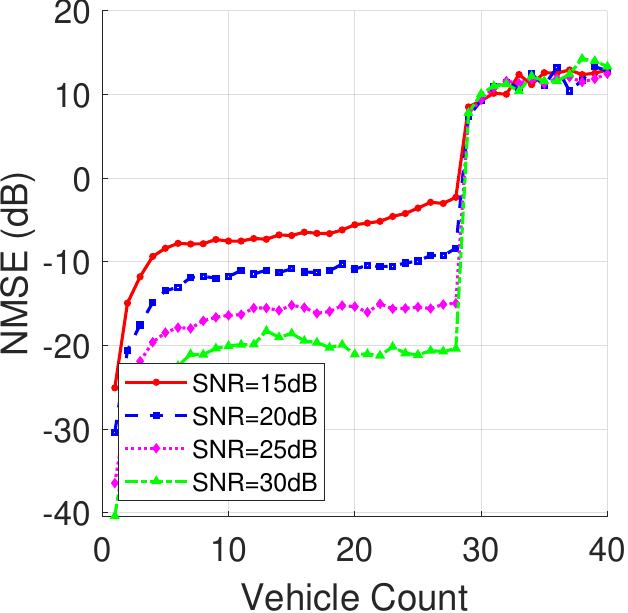}
    \caption{}
    \label{fig:sub_nmse}
  \end{subfigure}
  \hfill
  % 子图 (c)
  \begin{subfigure}[b]{0.3\linewidth}
    \includegraphics[width=\linewidth]{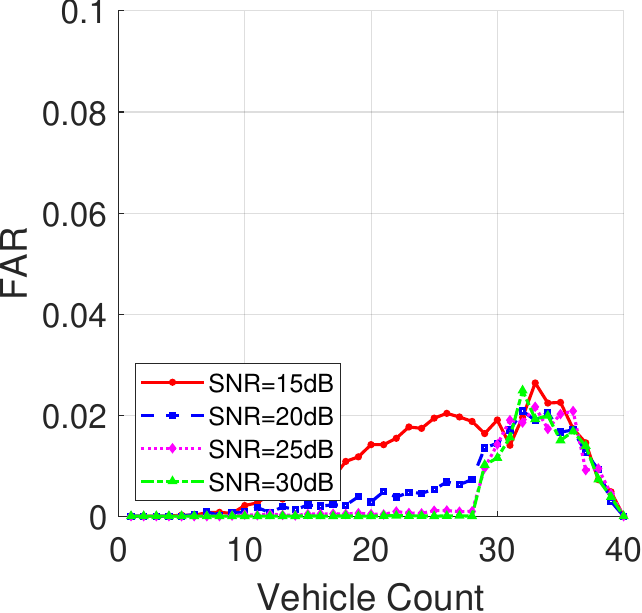}
    \caption{}
    \label{fig:sub_fdr}
  \end{subfigure}
  \hfill
  
  \caption{Parking lot surveillance performance versus the SNRs: (a) Detection Rate; (b) NMSE; (c) FAR.}
  \label{fig:combined_horizontal}
\end{figure*}

\begin{figure*}[!t] % 注意星号 *
  \centering
  % 子图 (a) - SLOT2
  \begin{subfigure}[b]{0.3\linewidth}
    \includegraphics[width=\linewidth]{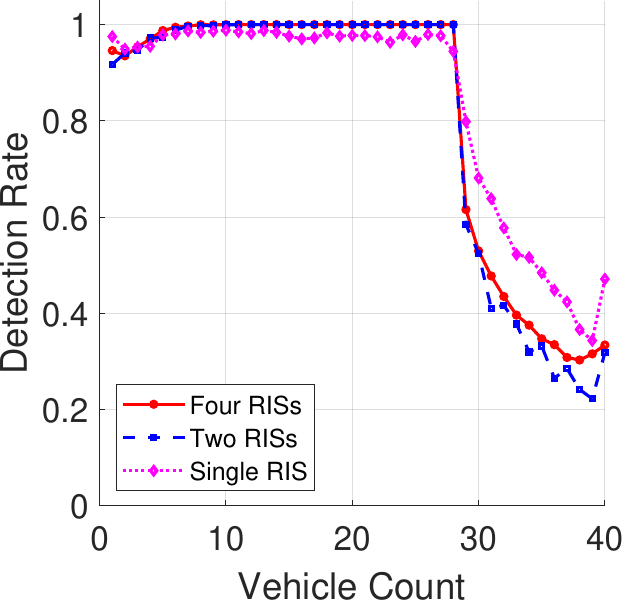}
    \caption{}
    \label{fig:sub_slot2}
  \end{subfigure}
  \hfill % 填充水平间距
  % 子图 (b) - SYSTEM2
  \begin{subfigure}[b]{0.3\linewidth}
    \includegraphics[width=\linewidth]{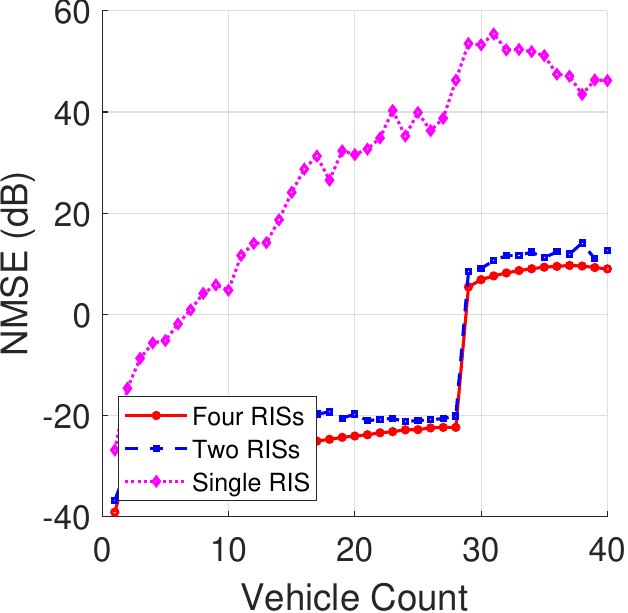}
    \caption{}
    \label{fig:sub_nmse2}
  \end{subfigure}
  \hfill
  % 子图 (c) - NMSE2
  \begin{subfigure}[b]{0.3\linewidth}
    \includegraphics[width=\linewidth]{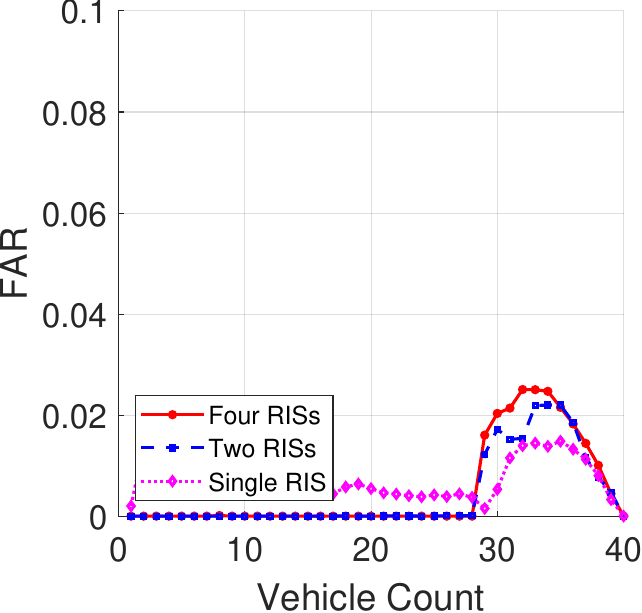}
    \caption{}
    \label{fig:sub_fdr2}
    \end{subfigure}
    \hfill
  
  \caption{Parking lot surveillance performance versus the number of RISs: (a) Detection Rate; (b) NMSE; (c) FAR.}
  \label{fig:combined_ris}
\end{figure*}

To evaluate the performance of the proposed method, we introduce three evaluation metrics. These metrics are briefly summarized as follows:
\subsubsection{Detection Rate} 
It is defined as:
\begin{equation}
  \text{Detection Rate} = D_1/D
  \label{eq:Accuracy},
\end{equation}
where \(D_1\) is the total number of correctly detected vehicles and \(D\) is the real number of vehicles in the parking lot.
\subsubsection{Normalized Mean Square Error (NMSE)}
It is defined as:
\begin{equation}
\text{NMSE} = \frac{1}{I} \sum_{i=1}^{I} \frac{\left\| \hat{\Delta\bm{\sigma}}^{(i)} - \Delta\bm{\sigma} \right\|_2^2}{\left\| \Delta\bm{\sigma} \right\|_2^2},
\end{equation}
where \( I = 1000 \) is the number of Monto Carlo simulations, and \( \hat{\Delta\bm{\sigma}}^{(i)} \) is the estimated image of \( \Delta\bm{\sigma} \) at the \( i \)-th simulation.
\subsubsection{False Alarm Rate (FAR)}
It is defined as:
\begin{equation}
  \text{FAR} = D_2/D
  \label{eq:FDR},
\end{equation}
 where \( D_2 \) represents the total number of false alarms, defined as the count of free parking spaces incorrectly classified as occupied.

\subsection{Simulation Results}
\subsubsection{Simulation with Different Signal-to-Noise Ratios (SNRs)}
The simulation results are shown in Fig.~\ref{fig:combined_horizontal}. We can find that for the same number of vehicles, the detection rate increases with increasing SNRs, while NMSE and FAR decrease, which suggests a positive correlation between the difference imaging performance of the parking lot and the SNRs. Next, we analyze the simulation results at the SNR of 30 dB as an example.

When the vehicle count is small, the detection rate generally increases with fluctuations as the number of parked vehicles increases, maintaining a high value of over 80\% (Figs.~\ref{fig:sub_slot}). When the vehicle count is in the range of \([10, 28]\), the detection rate achieves 100\%. However, it significantly decreases when the vehicle count exceeds 28. This indicates that the vehicle count is relatively large, causing the vector \(\mathrm{\Delta}\bm{\sigma}\) to be no longer sparse and significantly degrading the performance of the SP algorithm. When the vehicle count is less than 9, NMSE shows an increasing trend and reaches a plateau at -20dB until the count vehicle exceeds 28 (Figs.~\ref{fig:sub_nmse}). We can also find out that FAR no longer stays at 0, but fluctuates under 7\% when the vehicle count exceeds 28 (Figs.~\ref{fig:sub_fdr}).  

It is worth noting that when the vehicle count is small, the detection rate does not reach 100\% despite the very small NMSE (Figs.~\ref{fig:sub_slot},~\ref{fig:sub_nmse}). This is because NMSE reflects the global estimation precision, while the detection rate relies on the local threshold judgment. When the vehicle count is small, although the global error is tiny, the local misjudgment will be magnified, resulting in small loss in detection rate.

\subsubsection{Simulation with Different Number of RISs}

In this simulation experiment, we respectively utilize \( T = 1, 2, 4 \) RISs to achieve ROI imaging. The SNR is set at 30 dB. When there is merely one RIS, its location is at $[0, 0, 3]^{\text{T}}$ meters. For the scenario with four RISs, their locations are at $[7.5, 7.5, 3]^{\text{T}}$ meters, $[-7.5, 7.5, 3]^{\text{T}}$ meters, $[-7.5, -7.5, 3]^{\text{T}}$ meters, and $[7.5, -7.5, 3]^{\text{T}}$ meters, respectively.

The simulation results are shown in Fig.~\ref{fig:combined_ris}. We can find that for the same number of vehicles, the detection rate increases with increasing RIS numbers, while NMSE and FAR decrease, which suggests a positive correlation between the imaging performance of the parking lot and the number of RISs. Next, we analyze the simulation results with one RIS.

When the vehicle count is less than 7, the detection rate fluctuates and increases as the vehicle count rises, maintaining over 90\% (Figs.~\ref{fig:sub_slot2}). When the vehicle count is in the range of \([7, 27]\), the detection rate remains slightly lower than 100\%. However, the detection rate can be strengthened to 100\% with two or four RISs. When the vehicle count exceeds 27, the detection rate significantly decreases for all simulation scenarios. As the vehicle count rises, NMSE also continually rises and soon reaches above 0 dB, indicating poor imaging performance (Figs.~\ref{fig:sub_nmse2}). In general, the FAR with one RIS is much higher than that with two or four RISs, especially when the vehicle count is below 27 (Figs.~\ref{fig:sub_fdr2}). These results demonstrate that a single RIS may not accomplish the task of parking space surveillance while multi-RIS-aided collaborative difference imaging can achieve much better performance.

Moreover, it is worth noting that when the vehicle count is in the range of \([7, 27]\), the detection rate with one RIS can still remains considerably high while the NMSE is above 0 dB. The reason is as follows. When utilizing the SP algorithm to recover \(\mathrm{\Delta}\bm{\sigma}\), the absolute values of several elements in \(\hat{\Delta\bm{\sigma}}\) are wrongly estimated to be very large, making the NMSE large. However, by thresholding, we succeed in erasing the effect of these wrong estimations, and thus the detection rate is still quite high.

\section{Conclusion}\label{sec:vi}
This study proposes a difference imaging-based parking lot surveillance method in multi-RIS-aided collaborative ISAC system, addressing the limitations of traditional approaches and single-RIS configurations. By leveraging the sparsity of scattering coefficient changes between free and occupied status, we propose to transform the dense scattering coefficient recovery problem into a sparse difference image reconstruction task, enabling efficient CS-based solutions. The collaborative deployment of multiple RISs further enhances imaging performance. Simulation results demonstrate that the multi-RIS-aided system achieves 100\% detection rate in parking occupancy detection under moderate vehicle densities and sufficient SNR conditions. 

Our future research will focus on enhancing the robustness of the system in high-density parking scenarios by integrating prior knowledge or deep learning techniques into CS-based algorithms. We are also considering expanding the ROI into a 3D scenario for more accurate and specific detection. Additionally, RIS phase optimization strategies will also be explored to improve detection accuracy.

\section*{Acknowledgment}
This work was supported in part by the National Natural Science Foundation of China (NSFC) under Grant 62261160576 and Grant 624B2036, in part by the Fundamental Research Funds for the Central Universities under Grant 2242022k60004, in part by the National Science Foundation of Jiangsu Province under Grant BK20230818.

\end{document}